# PERCOLATION IN A MODEL TRANSIENT NETWORK: RHEOLOGY AND DYNAMIC LIGHT SCATTERING

E. Michel[+], M. Filali[++], R. Aznar[+], G. Porte[+], J. Appell[+] *

[+] Groupe de Dynamique des Phases Condensées

UMR 5581 C.N.R.S - Université Montpellier II , C.C.26

F-34095 Montpellier Cedex 05

[++]Laboratoire de Physique du Solide

Faculté des Sciences Dhar Mehraz

Fes Maroc

**Abstract:**

Step strain experiments and dynamic light scattering measurements are perfomed to characterize the dynamic behavior of an o/w droplet microemulsion into which is incorporated a telechelic polymer. At sufficient droplet and polymer concentrations, above the percolation threshold, the system is viscoelastic and its dynamic structure factor shows up two steps for the relaxation of concentration fluctuations: the fast one is dominated by the diffusion but the slower one is almost independent of the wave vector. The terminal time of the stress relaxation $\tau_R$ and the slow time of the dynamic structure factor $\tau_S$ are both presumably controlled by the residence time of a sticker in a droplet: consistently, $\tau_R$ and $\tau_S$ are of the same order, they both vanishes at the percolation threshold according to power laws but with different exponents. We discuss these features in terms of deviations at the transition, from the usual mean field description of the dynamics of transient networks.

*to whom correspondence should be addressed: **appell@gdpc.univ-montp2.fr**

**INTRODUCTION**

Quite a number of viscoelastic materials consist of a some kind of a transient network embedded within a liquid solvent. Hence, the dynamic light scattering is dominated by concentration fluctuations, the relaxation mechanism of which is more complex than the simple diffusion of usual brownian dispersions. At short times, the materials behaves like a solid and one observes a gel like mode. Usually this mode is overdamped due to the viscosity of the liquid solvent: it has a $q^2$ dependence revealing that it is dominated by mutual diffusion. At long time however, one observes in addition a q-independent mode revealing a second mechanism for which diffusion is not the rate limiting step. Indeed, the characteristic time of this second step is comparable to the terminal time in linear rheology. Theories reported to date [1-3] for the interplay between mutual diffusion and viscoelastic properties of such materials mainly address the case of semidilute polymer solutions for which a realistic quantitative description [4] of the linear rheology is available. On the other hand, a somewhat wider variety of viscoelastic materials has been investigated experimentally by dynamic light scattering (polymer solutions [5,6], giant micelles [7-10], thermoreversible aqueous copolymer gel [11], associating random block copolymer [12,13], networks of telechelic polymers [14-17]): two modes at least were clearly identified and even in some cases evidence of three modes was claimed.

Water soluble polymers with hydrophobic end blocks indeed exhibit viscoelastic behavior even at low concentration in binary aqueous solutions. These are indeed determined by the tendency for the end blocks to micellize into hydrophobic clusters linked together by the hydrosoluble threads and leading so to the formation of a reversible connected network. The rheological properties are controlled by the density of the network which determines the instantaneous elastic modulus and by the average residence time of a hydrophobic sticker in a given cluster which controls the characteristic time of the stress relaxation. Due to their many applications as commercial thickeners, these telechelic structures have been the subject of many publications [18-24] in the past ten years. Most studies however have focussed on pure polymer/water binary solutions although most practical applications involve other components amongst which surfactants are ubiquitous. Besides their practical interest, ternary mixtures, involving surfactant micelles and telechelic polymers in water, provides the possibility to control separately the average distance between clusters and the average degree of connections between clusters : the surfactant concentration monitors the number density of the micelles (the size of which are essentially controlled by the self assembling properties of the surfactant component) while the relative amount of telechelic polymers per micelle determines the connectivity of the network. Another advantage of ternary systems is that the contribution of the micelles to the scattering (light or neutron) heavily dominates that of the polymer, making so the analysis of scattering data easier and less ambiguous [25-29].

We investigate here the dynamic properties of such a quasi ternary system using quasi elastic light scattering and rheometry. The system consists of an oil in water droplet microemulsion with well controlled droplet size and shape into which moderate amounts of a hydrophobically end modified poly(ethelene oxide) PEO are added. Previous structural characterization [ 30] of this system using neutron scattering have shown that the size and shape of the droplets are not affected by the addition of the telechelic polymer. Moreover, the changes in the structure factor after addition of the polymer have revealed the tendency for the polymer to bridge neighbouring droplets. Consistently with this structural picture of droplets reversibly linked to one another, the samples exhibit strong viscoelastic behavior provided the polymer amount is large enough. We measure the time resolved stress response to moderate step strains using a strain controlled rheometer as function of the polymer concentration at fixed droplet concentration. And we compare these to the relaxation of the spontaneous fluctuations of the droplet concentration as measured by dynamic light scattering (DLS). Several characteristic features measured by these two techniques exhibit singular evolutions reminiscent of a percolation behavior: a finite minimum polymer concentration is required to built up an infinite transient connected network spanning the whole sample.

**EXPERIMENTAL SECTION.**

**Preparation of the samples.**

The droplet microemulsion involves two non-ionic surfactants : TX100 and TX35, the weight ratio $\Omega$ of which fixes the spontaneous curvature of the surfactant film. The surfactant micelles are swollen with decane at a constant weight ratio $\Gamma$ with the surfactant ($\Gamma$ = [decane]/[TX100+TX35]). Appropriate choice for $\Omega$ and $\Gamma$ provides spherical droplets with the desired radius : we used $\Omega$= 0.5 and $\Gamma$= 0.7 . Neutron scattering (figure 1) revealed 82Å spherical droplets with low polydispersity: their size and shape remain constant upon variation of the concentration in droplets and of the amount of added polymer.[30]

The poly (ethylene-oxide) polymer has been hydrophobically modified and purified in the laboratory using the method described in [31,32]. The molecular weight of the starting products is determined by size-exclusion chromatography. After modification, the degree of substitution of the hydroxyl groups was determined by NMR using the method described in [33]. The degree of substitution is found to be equal or larger than 98% . The hydrophobically modified poly (ethylene-oxide)-PEO-2M contains an isocyanate group between the alkyl chain $C_{18}H_{25}$ and the ethylene-oxide chain.

The aqueous solutions are characterized by their volume fraction $\Phi$ of aliphatic chains from decane, TX and PEO-2M which form the hydrophobic cores of the microemulsion droplets, and by the number r of $C_{18}$ chains per droplet. They are prepared by weight in triply distilled water in order to obtain samples with given $\Phi$ and r. All the parameters necessary to calculate $\Phi$ and r from the sample composition are summarized in Table 1; r is calculated assuming the radius of the droplets is equal to 82Å [30] .

**Phase behavior.**

Since the size and shape of the droplets are essentially invariant, we consider the system as a quasi ternary system (droplets/polymer/water) and we represent in figure 2 the phase behavior (at fixed pressure and temperature) as function of two variables : the droplet concentration $\Phi$ (volume fraction of the hydrophobic cores of the droplets) on the horizontal axis and the number r of hydrophobic stickers per droplet (twice the number of polymers per droplet) on the vertical axis. A phase separation is seen at high r and low $\Phi$: neutron scattering data (not shown here) taken from the upper and lower phase respectively reveal that both consist of droplets of the same radius (82Å) but with different concentrations. The critical point associated to this liquid-gas phase separation is approximately located by examination of the strong turbidity of the samples. Note that the critical point does not coincide with the lower point on the coexistence curve which means that the binodal are not horizontal but tilted: the upper phase, more concentrated in droplets, is also more concentrated in polymer than the lower more dilute phase. Of course this phase behavior is controlled by the effective attractive contribution arising from the bridging between droplets [ 20, 34-36]. The "percolation" line tentatively sketched in the lower part of the diagram is drawn according to the data reported in later sections.

**Dynamic behavior of a generic viscoelastic sample**

<u>1) Rheometry: step strain experiments</u>**.** All rheological data presented here are obtained from step strain experiments using a Rheometrics RFS II strain controlled rheometer in the cone and plate geometry. At time t =0, the sample is abruptly submitted to a sudden step shear strain of amplitude γ. And the time resolved shear stress response σ(t) is recorded : the time resolution of the rheometer is of the order of 0.02s ; so we discard data points recorded before time t =0.05s. We here restrict our attention to the linear regime for which :

$$G(t) = \sigma(t) / \gamma \tag{1}$$

does not depend on γ. Comparing stress responses after increasing step strains, we checked that the linear regime extend up to γ =0.3 at least for all samples investigated here. Figure 3 shows a typical stress relaxation : it is taken from the sample Φ =12.4 % and r =18 after a step strain of γ =20%. We see an almost linear decay in lin/log units which indicate an almost Maxwellian behavior. However a better fit is obtained using a slightly stretched exponential with an exponent of 0.82 :

$$G(t) = G(0) \exp[-(t/\tau_R)^{0.82}] \tag{2}$$

with G(0) = 1830 Pa and $\tau_R$ = 0.125 s. Note that $\tau_R$ is not large compared to the time resolution of the rheometer : quite an important initial portion of the stress relaxation is therefore missed due to the lack of resolution of the rheometer. However a very similar stress relaxation pattern (same exponent for the stretched exponential) have been observed for a similar system but with a different surfactant exhibiting relaxation times of the order of 1s : the fit could therefore be extended down to much lower values of t / $\tau_R$ for this other system. We could check that the values for G(0) and $\tau_R$ extracted from the same fitting procedure was quite insensitive to the number of experimental points discarded in the initial decay up to values of t /$\tau_R$ of the order of 1. We therefore keep confidence in the here obtained values of G(0) and $\tau_R$. Right in line with the usual interpretation in terms of the reversible network theory, we expect the stress relaxation to arise from the finite residence time $\tau_{Rt}$ of a sticker in a given droplet. Since the escape of a given sticker from a droplet is presumably a thermally activated process, we expect $\tau_{Rt}$ and therefore $\tau_R$ to exhibit an Arrhenius dependence versus the temperature:

$$\tau_R = \tau_0 \exp(E_S/k_B T) \tag{3}$$

where $\tau_0$ is some inverse frequency of attempts and the activation energy $E_S$ is the reversible work of extraction of the sticker from the hydrophobic core into the free water. Figure 4 shows the evolution of $\tau_R$ as function of the inverse temperature for the same sample as above. The Arrhenius dependence indicate an activation energy $E_S = 43\ k_B T$. We shall see however that $\tau_R$ does not simply coincide with the life time $\tau_{Rt}$.

2. Rheometry: steady shear measurements. Submitting the same sample to steady shears at increasing rates $\dot{\gamma}$ with the same rheometer (figure 5) we obtain a shear thinning flow curve with a discontinuous instability at high shear rate: $\dot{\gamma} \geq \tau_R^{-1}$. Systematic measurements with shear cells of different geometries clearly indicate that this instability takes place close to the walls of the shearing cell, however the signature is not that of simple sliding ar the walls. We are currently investigating this instability which we shall discuss in a forthcoming article. At low shear rate however, a Newtonian behavior is recovered with a well defined low shear viscosity: $\eta(\dot{\gamma} \to 0)$. For the above sample we find $\eta(\dot{\gamma} \to 0) = 195$ Pa. s$^{-1}$. Note that, consistently with the quasi-Maxwellian stress relaxation the low shear viscosity is of the same order as the product of the modulus $G(0)$ by the terminal relaxation time $\tau_R$: $\eta(\dot{\gamma} \to 0) \sim G(0) \cdot \tau_R \sim 228$ Pa. s$^{-1}$.

3-Dynamic light scattering. The measurements are performed on a standard setup (AMTEC Goniometer + Brookhaven correlator) the light source is an argon ion laser ($\lambda = 514.5$ nm). The full homodyne intensity autocorrelation function was measured at different q's ; the wave vector $q = \frac{4\pi n}{\lambda} \sin\frac{\theta}{2}$ with $\lambda$ the wavelength, $\theta$ the scattering angle and n the refractive index of the microemulsion; q range from $3\ 10^{-4}$ to $3\ 10^{-3}$ Å$^{-1}$. If the scattered field obeys gaussian statistics the normalized autocorrelation function $g_2$ (q,t) :

$$g_2(q,t) = \langle I_q(t) I_q(0) \rangle / \langle I_q \rangle^2 \qquad (4)$$

can be related to the theoretically amenable first-order electric field correlation function $g_1$ (q,t) by the Siegert relationship: $g_2(q,t) - 1 = |g_1(q,t)|^2$. In our system, the scattered light mostly arises from the droplets and the contribution of the polymer is negligible. The measurement therefore probes how fluctuations of the concentration of the droplets relax at a length scale 1/q: the time dependence of $g_1$ (q,t) is thus given by :

$$g_1(q,t) = \langle \delta\Phi_q(t) \delta\Phi_{-q}(0) \rangle / \langle |\delta\Phi_q|^2 \rangle. \qquad (5)$$

The normalized intensity autocorrelation function $g_2(q,t)-1$ taken at $\theta=90$ (q= $2.3\ 10^{-3}$ Å$^{-1}$) for the same sample as above $\Phi=12.4\%$ and r =18 is shown in figure 6A; for the sake of comparison we show in figure 6B the same pattern obtained for a microemulsion with the same concentration but no added polymer ($\Phi$ =12.4%, r=0). For the bare microemulsion, we find a single exponential relaxation (full line in figure 6B) driven by the collective diffusions of the droplets. This is confirmed by the q-dependence (data not shown) of the characteristic time: $\tau^{-1} = D_{coll}\ q^2$, where $D_{coll}$= $2.9\ 10^{-11}$ m$^2$ s$^{-1}$ , a value consistent with the radius for the droplets as measured by neutron scattering. The pattern in figure 6A is different indicating a relaxation in two steps at least, with very different time scales: the first step has a characteristic time comparable to the diffusion in the bare microemulsion whereas the second one is thousands time slower with a time scale comparable to that of the stress relaxation in the above step strain experiment. Accordingly, the pattern was fitted (full line in figure 6A) with the expression:

$$g_2(t) - 1 = \left[ A_f \exp(-(t/\tau_f)) + A_s \exp\left(-(t/\tau_s)^{0.82}\right) \right]^2 \qquad (6)$$

where the indices s and f stand for the slow and the fast contribution respectively. Actually the quality of the fit is very good. The q-dependence of $\tau_f$ in figure 7: ($\tau_f \sim q^{-2}$) indicates a simple collective diffusion process with an effective diffusion coefficient $D_{coll} = 3.5 \; 10^{-11} \; m^2 \; s^{-1}$ close to but slightly larger than that of the bare microemulsion. On the other hand, $\tau_s$ in figure 8 appears to depend weakly on q : the slow step is thus not limited by diffusion. The relative amplitudes of the two relaxation mechanisms (cf figure 8) shows only a weak dependence on q. For the sake of consistency we derive from those plots $\tau_{s0}$ and $[A_s/(A_s+A_f)]_0$, the values extrapolated to zero wave vector.

**Evolution with r : evidence for percolation**

We investigate systematically the dynamic behavior as function of r at fixed droplet concentration: $\Phi = 12.4 \; \%$. This particular concentration was chosen because it corresponds to an average distance between droplets comparable to the end to end distance in a gaussian coil having the same full length as the polymer: in such conditions, the polymer can easily bridge neighbouring droplets without being forced to stretch

<u>1) Rheometry</u>. For all samples G(t) could be exploited following the procedure described in the above subsection, we use the same stretched exponential for the fit (same exponent 0.82) although close to the percolation threshold the agreement is not so good as at higher r. G(0) and $\tau_R$ are plotted versus r in figure 9A and 9B whereas the variations of the viscosity $\eta$ ($\dot{\gamma}$->0) are shown in figure 10. All three quantities decrease with decreasing r and vanish below a finite value of r of the order of 4, suggesting strongly a percolation process. Accordingly [37], we fitted the evolution of the three quantities according to power laws of the form $(r - r_p)^\beta$ where $r_p$ is the number of stickers per droplet at the percolation point. Good fits were obtained (full lines in figure 9A, 9B and 10) for the three quantities with the same value for $r_p = 4.23$, but different exponents $\beta$'s:

$$\eta(\dot{\gamma}->0) = 0.89(r-4.23)^{2.06} \text{ Pa.s} \quad (7a)$$

$$G(0) = 45(r-4.23)^{1.42} \quad \text{Pa} \quad (7b)$$

$$\tau_R = 0.032(r-4.23)^{0.49} \quad \text{s} \quad (7c)$$

Note again that consistently with the quasi Maxwellian behavior of all samples, the exponent for $\eta$ ($\dot{\gamma}$->0) is close to the sum of the exponents of G(0) and of $\tau_R$.

2) Dynamic light scattering. The correlation functions $g_2(t)-1$ at various r are shown together on figure 11. Clearly the relative contribution of the slow mode decreases and vanishes when approaching the percolation point; correlatively, the relaxation time decreases as well. On figure 12A and 12B we plotted respectively the evolutions of $[A_s / (A_f + A_s)]_0$ and of $\tau_{S0}$ versus r. On the same figures, we force fitted these quantities with power laws where the percolation connectivity index is fixed at the same value as derived from the above step strain measurements: $r_p = 4.23$. Actually the quantitative agreement is good, again supporting strongly the idea that the percolation is indeed relevent for the slow mode of concentration fluctuations as well as for the stress relaxation. The effective exponents extracted from these fits are:

$$\left[A_s/(A_f + A_s)\right]_0 = 0.26 \, (r - 4.23)^{0.37} \tag{8a}$$

$$\tau_{s0} = 0.06(r - 4.23)^{0.9} \tag{8b}$$

We notice that the stress relaxation time $\tau_R$ and the slow time $\tau_S$ in DLS not only differ in magnitude but have different evolutions –different exponents- as function of the density of links.

On the other hand (figure 13), the effective mutual diffusion coefficient, as derived from the fast mode analysis, only shows a mild dependence on r with no singularity at the percolation point: the initial fast step is therefore hardly affected by the connectivity of the transient network.

**DISCUSSION**

As a main finding of our experiments, several dynamic characteristics vanish at a finite degree of connectivity suggesting percolation: the instantaneous shear modulus $G(0)$, the terminal relaxation time $\tau_R$, the low shear viscosity $\eta \, (\dot{\gamma} \to 0)$, and the relative amplitude $[A_s / (A_f + A_s)]_0$ and the characteristic time $\tau_{S0}$ of the slow step in dynamic light scattering. (Others do not: the collective diffusion coefficient $D_{coll}$ as well as the amplitude of the fast mode in dynamic light scattering show no singularity). We underlined the singular evolutions at the percolation by fitting the corresponding data with power laws and the question arises of the accuracy of the exponents that we determine accordingly. The situation is not very good for $G(0)$ and $\tau_R$. As noted above, the time resolution of the rheometer (0.05s) is a serious limitation specially close to the percolation where the stress relaxation time $\tau_R$ is short. Values extracted by extrapolation to $t = 0$ may well be misestimated. In dynamic light scattering, the contribution of the slow process vanishes at the percolation, so the accuracy for the relative amplitude and for the characteristic time is not very good either. The situation is much better for the low shear viscosity $\eta$ ($\dot{\gamma} \to 0$) for which the accuracy remains very good even for the most fluid samples. This is the reason why in practice we proceeded in the following manner. We first fitted the low shear viscosity data for which the accuracy is best even close to the threshold; we so obtained reliable estimations of both $\beta_{\eta\,(\dot{\gamma} \to 0)}$ and $r_p$. Then we fitted all the other quantities forcing the threshold $r_p$ at the same value and leaving the exponent as the only free adjustable parameter. Actually, the agreement looks good. But, if we consider each quantities separately, we can indeed obtain equally good fits with different thresholds and exponents: the value of the exponent then depends very sensitively on the value taken for the threshold. At the present time, we have no firm theoretical grounds to support the assumption of one common threshold for all quantities. We mention in the next paragraph the difference between geometrical connectivity percolation (likely relevent for DLS) and structural rigidity percolation (likely relevent for rheology). So, we certainly cannot claim for an accurate quantitative characterization of the exponents at the threshold. Another intriguing aspect of our data deserves to be underlined. In principle, in percolation situations, the singular power law dominates the evolution of a given quantity only close to the threshold. Far above, the mean field behavior is usually recovered. In contrast, our fits happen to be surprisingly good even far above $r_p$, a feature that we do not understand. This is a further reason to be very cautious with the exponents we determine from such fits which include data taken far above $r_p$. We discuss the behavior of each quantity in the following paragraphs.

$G(0)$, as measured in a step strain experiment, characterizes the immediate elastic response of the network to a sudden deformation, before any relaxation due to the finite life time of a link. It is natural

---

in this picture that G(0) vanishes below a finite value $r_p$ of the connectivity parameter: below $r_p$ there is no crosslinked infinite path, connecting continuously the cone and the plate, and capable to sustain the transient elastic torque. Actually, data reported in the literature for situations similar to ours displayed essentially the same features and were analysed accordingly. Bagger-Jörgensen et al [26] found an exponent $\beta_{G(0)} \approx 1.5$ and their percolation threshold was about 2.1 as expressed in number of polymer per droplet: i-e 4.2 in terms of number of stickers per droplet. These values are remarkably close to ours: their experimental conditions (average distance between droplets and polymer molecular weight) were in fact also very close to ours. The data of Schwab and Stühn [17] display a percolation behaviour for G(0) as well but with a slightly different exponent: $\beta_{G(0)} \approx 1.7$. Since by definition G(0) does not involve any feature related to relaxations, its evolution can be compared to theoretical predictions derived at true sol-gel transitions [37]. The exponent calculated for the elastic modulus is 1.7 above the gel point: this is very close to the value of Schwab and Stühn but higher than ours. However, as underlined above, one should not overestimate the accuracy of the fits. The value we find for the threshold $r_p = 4.2$ is higher that expected from simulation on a cubic lattice: $\approx 1.5$ [38]. Part of the discrepancy arises from the number of threads that loop on the same droplet and therefore do not participate to active links. On the other hand, the value predicted from simulations is derived assuming that one bond at most can link two neighbouring sites whereas in our experimental situation, two or even more chains may link two adjacent droplets. Moreover, polymer links only sustain central forces between droplets and do not oppose any resistance to reorientations at fixed length. Then, elasticity requires not only an infinite path of links connecting the cone and the plate but also that this path is sufficiently triangulated so that the stress is not immediately released by simple bond reorientations.

---

Michel et al                                  Juin 2000                                            14

This leads to the notion of a "rigidity" threshold which is certainly somewhat higher than the usual "connectivity" threshold [39].

The terminal relaxation time $\tau_R$ is related to the residence time of a sticker in a droplet. In the usual interpretation of the stress relaxation in transient networks [40,41] the spatial distribution of the nodes is assumed to be affinely deformed by the step strain and the length distribution of the links is thus shifted accordingly. The transient off equilibrium length distribution is at the origin of the measured stress. From time to time, stretched links disengage due to the finite residence time of their stickers and reconnect with the equilibrium length distribution: they "forget" the initially imposed strain and no more contribute to the stress. In this picture the stress at time t is a simple measure of the number of links that still reminds the initial strain after time t and we would expect $\tau_R$ to be simply identical to the residence time. Our measurements do not support this expectation. We find that $\tau_R$ sensitively depends on the average degree of connectivity r and vanishes at $r_p$. Whereas the residence time is completely determined by the adsorbtion energy of a sticker in a droplet. It should not depend on non local features such as the degree of connectivity of the network. To understand the discrepancy, we note that the above affine picture is a mean field description which assumes that the imposed strain distribute homogeneously within the network. Such homogeneity certainly breaks when approaching the percolation point. Close to the threshold, the infinite connected cluster consists of more densely crosslinked subclusters connected to each other by weaker parts where the links are less dense. Breaking a small number of links only, in a weak part, will suddenly release the stress within the whole adjacent dense subclusters. In this non mean field picture, we expect $\tau_R$ to be shorter than the residence time and indeed to vanish at the percolation as observed in the experiment.

---

When approaching a true sol-gel transition from below, the viscosity of the sol increases with the degree of connectivity and diverges at the transition. Above the transition, the viscous behavior of the sol is replaced by the elastic behaviour of the gel. Although we have not measured systematically the viscosity below $r_p$, we indeed expect it to exhibit a similar behavior related to the divergence of the size of the connected clusters at the percolation. Above a true gel transition, speaking of viscosity makes no sense because a gel is elastic and does not flow under mild enough stresses. In our system however, the terminal time $\tau_R$ is finite and the material ultimately flows beyond $\tau_R$ whatever low is the applied strain: the notion of low shear viscosity $\eta\ (\dot{\gamma} \to 0)$ makes sense above the percolation point. Just like the instantaneous elastic modulus $G(0)$, it is dominated above $r_p$ by the behavior of the infinite cluster. So we expect it to decrease in a singular manner when approaching the percolation point from above in agreement with our measurements. Consistently with the quasi Maxwellian stress relaxation at all r, we find: $\eta\ (\dot{\gamma} \to 0) \approx G(0)\ \tau_R$, and the exponent of the viscosity is close to the sum of that of the elastic modulus and that of the terminal time ($\beta_{\eta\ (\dot{\gamma} \to 0)} \approx \beta_{G(0)} + \beta_{\tau_R}$)

The current litterature dealing with dynamic light scattering on transient networks is more confusing. At least two and even sometimes three steps are reported for the relaxation of concentration fluctuations. In some cases[14-16] all three steps are dominated by diffusion ($\tau^{-1} \propto q^2$); in other reports [13, 17], the longest and the shortest steps are diffusion like whereas the intermediate step is q-independent. In order to get rid of possible artifacts, we spent a lot of time caring about the quality of the samples. In some early measurements, we also found a third very slow relaxation in addition to the two steps described in the experimental section. The samples however appeared slightly misty due to a

---

residual amount of insoluble parrafin in the polymer batch. Further purification of the polymer suppressed the misty aspect. And with the purified samples, all our attempts to detect a third relaxation failed. The absence of the third relaxation is somewhat puzzling. A consistent coarse grained description of the local state in the sample involves at least three distinct internal variables: the droplet concentration $\Phi$, the active links to dead loops ratio $\alpha$, and the number of polymer per micelle r/2. Indeed, DLS only probes the relaxations of $\Phi$, but $\Phi$ is coupled to both $\alpha$ and r. $\Phi$ and r are conserved variables (conservation of the total amount of droplets and of polymer), whereas $\alpha$ is not conserved (changing locally the link to loop ratio does not imply transport of matter). We therefore expect three dynamic modes: two of them being dominated by diffusion and the remaining one being essentially independent of q (if the caracteristic timescales are sufficiently different).

In practice, our DLS results can be qualitatively interpreted as follows. The short time, diffusive regime corresponds to relaxation of concentration fluctuations at fixed connectivity for each droplet. The corresponding diffusion coefficient is essentially comparable to that of the bare microemulsion but slightly accelerated by the spring like elasticity of the links. This diffusive motion then saturates after some times reflecting the fact that part of the concentration fluctuations arises from fluctuations of the connectivity density which is frozen at short time. This accounts for the shoulder in the dynamic structure factor. This intermediate plateau would last for ever if the system was in a true gel state with infinite life time for the links. But the system is liquid and the connectivity is renewed completely after delays of the order of the residence time: the relaxation of the concentration fluctuation then proceeds to completeness. This second step still involves diffusion but the rate limiting step is not the motion of the free droplets but rather the renewal of the connectivity. Since, as analysed above, the connectivity $\alpha$ is not conserved, we expect the characteristic time to be q-independent and indeed the slow time $\tau_S$ is only weakly q-dependent. Following this analysis, $\tau_{S0}$ should compare with the terminal time of the rheology $\tau_R$ which also corresponds to the renewal of the connectivity. However, although both times vanish at the percolation threshold, they not only differ up to some prefactor but clearly show different exponents when approaching the percolation. And the question arises of the origin of this difference in behavior?

Actually, the above qualitative analysis of the dynamic structure factor would apply to any binary viscoelastic material [2]. Comparison with the behavior of another viscoelastic system, but having a very different structure, suggests that the difference between $\tau_{S0}$ and $\tau_R$ is specific to our system. Semi-dilute solutions of wormlike micelles are known to exhibit a two step dynamic structure factor [9,10]. We prepared a semi-dilute solution of wormlike micelles by dissolution of cetylpyridinium chloride and Na-salicylate (mole ratio NaSal/CPCl=0.6) in a 0.5M NaCl aqueous solution: the total weight fraction of CPCl and salicylate was 12 %. At such a concentration, the solution shows up a strong viscoelastic behavior as revealed by the quasi-Maxwellian stress relaxation shown in figure 14A. It indeed arises from the entanglements between the wormlike micelles so deep into the semi-dilute regime. The dynamic structure factor shows two relaxation steps: the shorter indeed has the expected q-dependence for diffusion whereas the longer one is remarkably independent of q. We plot on figure 14B the slow step in DLS of a typical semi-dilute solution of giant wormlike micelles. Interestingly, the terminal time $\tau_R$ of the rheology ($\tau_R = 0.52$ s) and the characteristic time $\tau_S$ of the slow step in the dynamic stucture factor ($\tau_S = 0.5$ s) are almost identical. We checked that this coincidence actually resists to change in the experimental conditions (concentrations and temperature).

---

So there is clearly an unexpected specific feature in our transiently linked droplets. It may originate in the geometry of the phase diagram in figure 2. The two phase coexistence implies the existence of a spinodal line tangent to the binodal line at the critical point. So moving vertically in the phase diagram, increasing r above the percolation threshold at fixed $\Phi = 12.4$ %, one gets closer and closer to the spinodal line. The vicinity to this line should not affect the stress relaxation which only probes the connectivity renewal. But it may bring an additionnal slowing down for the slow step of the relaxation of concentration fluctuations for which the ultimate driving force is the inverse osmotic compressibility (at relaxed connectivity) which goes to zero at the spinodal line. To check further this point, we did the same rheology and QELS measurement for samples along a dilution line at fixed connectivity index r=18 in the direction of the critical point $P_C$. $\tau_R$ decreases upon dilution consistently with the expectation that at fixed r the relative amount of active bridges (versus dead loops) indeed decreases upon dilution. On the other hand $\tau_{S0}$ just shows the opposite trend: it increases although the density of active bridge decreases. This means that the slowing down due to the increasing osmotic compressibility overcompensates the accelerating effect due to the decreasing amount of active connections.

To sum up briefly, we have investigated the dynamic properties of a model transient network using rheometry and dynamic light scattering. Several characteristic features vanish below a well defined, finite minimum value of the relative concentration in telechelic polymer suggesting strongly a percolation behavior. This is indeed not unexpected since a minimum amount of polymer is certainly required to build up a reversible network spanning the whole sample and thus capable either to sustain transiently an elastic stress or to limit for some time the complete release of a concentration fluctuation. But beyond this qualitative picture, some intriguing facts appear in our data. First of all, power laws still fit the data far above the threshold: the mean field behavior is still not recovered at a polymer concentration five times larger than that of the threshold. Second, the absence of a third very slow step in DLS suggests that the polymer concentration is only weakly coupled to the droplet concentration. Finally, the different evolutions of the stress relaxation time and the characteristic time of the slow step in DLS probably arise from the vicinity of the phase separation determined by the attractive bridging interaction between the droplets.

**Acknowledgements.**

One of us, R.A., is grateful to Michel Viguier and André Collet for their precious advices during the synthesis of PEO-2M.

Table 1 Molar Mass and density of the components of the samples

| Component (abreviated in the text) | Molar Mass (dalton) | | Density (g/cm$^3$) | |
|---|---|---|---|---|
| | | HC(a) | polar part | HC(a) |
| $H_2O$ | 18 | - | 1 | |
| $[H_3C-(C-(CH_3)_2-CH_2-C-(CH_3)_2)\varphi]$ $(O-CH_2-CH_2)_{9.5}$ -OH (TX100) | 624 | 189 | 1.2 | 0.86 |
| $[H_3C-(C-(CH_3)_2-CH_2-C-(CH_3)_2)\varphi ](O-CH_2-CH_2)_3$ -OH (TX35) | 338 | 189 | 1.2 | 0.86 |
| $[H_3C-(CH_2)_8 CH_3]$ (decane) | 142 | 142 | - | 0.75 |
| $[CH_3-(CH_2)_{11}]-NH-CO-(O-CH_2-CH_2)_{227}$ | ~11 000 | 506 | 1.2 | 0.81 |
| -O-(CO)-NH-[ $(CH_2)_{17}$ $CH_3$ ]    (PEO-2M) | | | | |

a) HC = hydrophobic part of the component indicated in brackets in column 1

**FIGURE CAPTIONS**

Figure 1 A small angle neutron scattering pattern in the Porod representation for the microemulsion droplets prepared in deuterated water. The open circles are the data points for a sample with a volume fraction of hydrophobic cores $\Phi = 0.014$. The solid line correspond to the form factor of spherical droplets with a gaussian distribution of size with a mean radius of 82Å and a standard deviation of 7Å .For more details see ref [30]

Figure 2 Phase behavior of the quasi ternary system (droplets/ PEO-2M / water) as a function of the two variables: $\Phi$ the volume fraction of the hydrophobic cores of the droplets and r the number of hydrophobic stickers (C18) per droplet: r = twice the number of polymers per droplet. $P_c$ is the critical point associated to the phase separation observed at low $\Phi$ and high r values (see text). The dotted line is a tentative drawing of the percolation line derived from preliminary step-strain measurements. The samples studied here lie on the thin vertical line with $\Phi=12.4\%$.

Figure 3 Stress relaxation curve after a step strain of $\gamma= 20\%$ for the sample $\Phi = 12.4\%$ and r= 18. The solid line is the fit of the experimental data with the expression given by (2) with $G(0) =1830$ Pa and $\tau_R =0.125$ s

Figure 4 Arrhenius plot: $\text{Log}\tau_R= f(1/T)$ for the sample $\Phi = 12.4\%$ and r= 18. From the straight line we obtain using relation (3) $E_S \sim 43\ k_BT$ .

Figure 5 The flow curve = viscosity $\eta$ as a function of the rate $\dot\gamma$ for the sample $\Phi = 12.4\%$ and r= 18. $\eta (\dot\gamma -> 0) = 195$ Pa.s. Note the abrupt drop in viscosity around $2s^{-1}$ .

Figure 6 Normalized autocorrelation curve measured at $\theta = 90°$ ($q= 2.3\ 10^{-3}\ \text{Å}^{-1}$).Figure 6A: for the sample with $\Phi = 12.4\%$ and r= 18. The solid line is the fit of the data with relation (6) and the parameters $A_f = 0.29$, $\tau_f = 46\ \mu s$ and $A_s= 0.56$, $\tau_s = 0.6$ s. Figure 6B for the corresponding bare microemulsion $\Phi = 12.4\%$ and r= 0 . The solid line is a fit to a single exponential with $\tau = 64\ \mu s$

---

Figure 7  For the sample with $\Phi$ = 12.4 % and r= 18, illustration of the variation of $1/\tau_f$ with $q^2$. The slope of the line yields $D_{coll}$ = 3.5 $10^{-11}$ $m^2$ $s^{-1}$ .

Figure 8  For the sample with $\Phi$ = 12.4 % and r= 18, illustration of the weak dependence on q of $\tau_S$ figure 8A and of $A_S / (A_f + A_S)$ figure 8B

Figure 9  For samples with $\Phi$ = 12.4 % : Evolution with r of G(0) and $\tau_R$ derived from the experimental results as described in figure 3 for the sample with r=18.  Figure 9A: G(0), the solid line is the fit of the data with a power law given by relation (7b). Figure 9B : $\tau_R$ the solid line is the fit of the data with a power law given by relation (7c).

Figure 10  For samples with $\Phi$ = 12.4 % : Evolution with r of the viscosity $\eta$ ($\dot{\gamma}$ ->0) derived from the experimental results as described in figure 3 for the sample with r=18. The solid line is the fit of the data with a power law given by relation (7a).

Figure 11  The normalized intensity autocorrelation curves measured at $\theta$ = 90° (q= 2.3 $10^{-3}$ $Å^{-1}$) for samples with $\Phi$ = 12.4 % and r= 0,3,6,9,12,15,18,21 for the curves from left to right.  The solid line through the experimental points are the fits calculated using relation (6).

Figure 12  Evolution with r of the relative amplitude $[A_S / (A_f + A_S)]_0$ and of the relaxation time $\tau_{S0}$ for the slow relaxation mode derived from the curves of figure 11. 12A : $[A_S / (A_f + A_S)]_0$ the solid line is the fit of the data with a power law given by relation (8a) and 12B : $\tau_{S0}$ . the solid line is the fit of the data with a power law given by relation (8b).

Figure 13  Evolution with r of the collective diffusion coefficient $D_{coll}$ corresponding to the fast relaxation time from the curves of figure 11.

Figure 14  Dynamical behavior of another type of transient network : a semi -dilute solution of wormlike micelles : Cetyl pyridinium chloride + salicylate of sodium, weight fraction of micelles = 12%. 14A: The stress relaxation curve G(t) for the same sample is an exponential with $\tau_R$ = 0.52s. 14B:

The second relaxation mode in dynamic light scattering is described by an exponential decrease with $\tau_S$ = 0.5s.

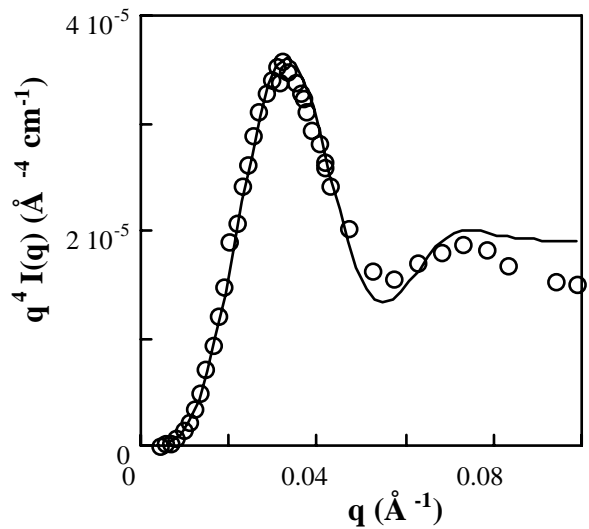

Michel et al   Figure 1

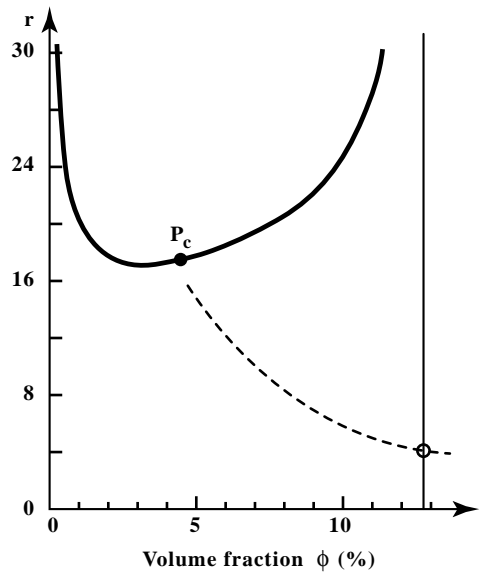

Michel et al   Figure 2

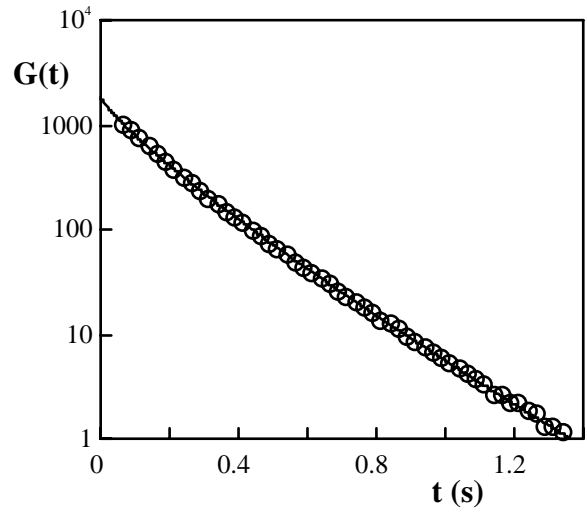

Michel et al  Figure 3

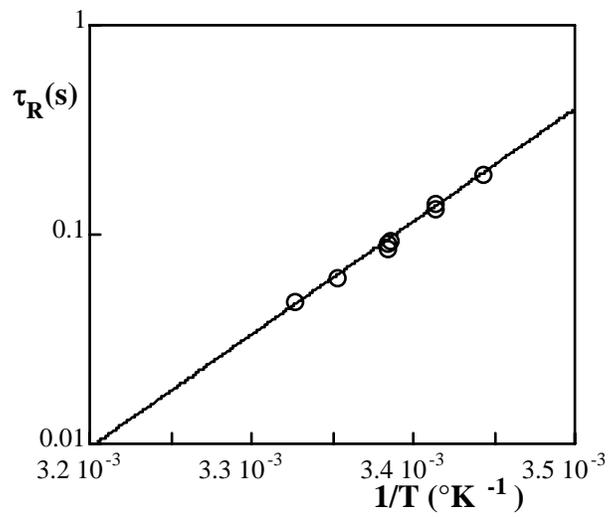

Michel et al  Figure 4

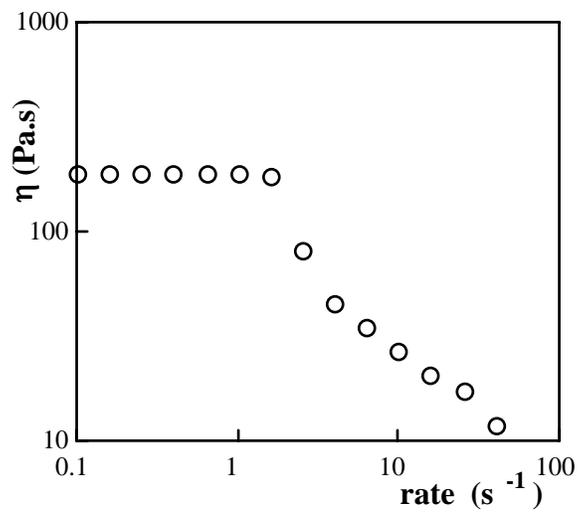

Michel et al  Figure 5

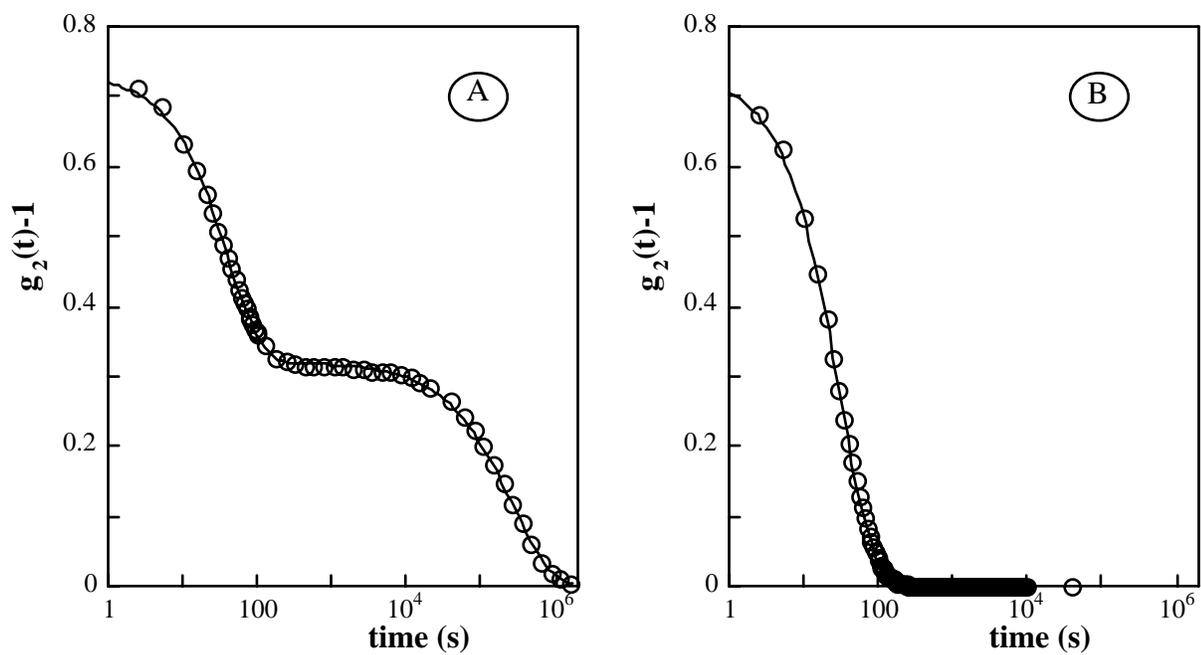

Michel et al  Figure 6

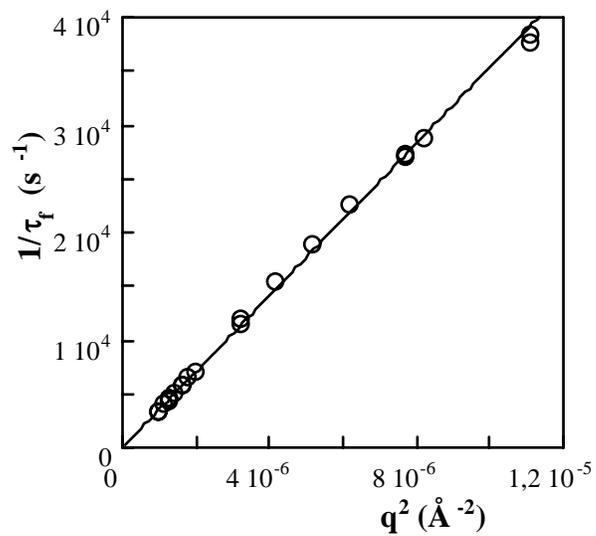

Michel et al   Figure 7

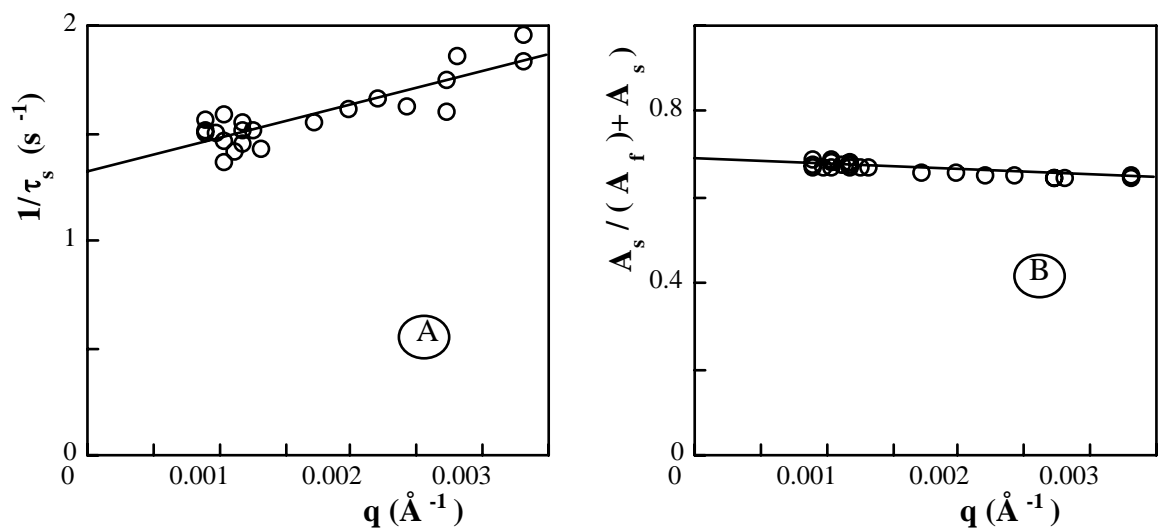

Michel et al   Figure 8

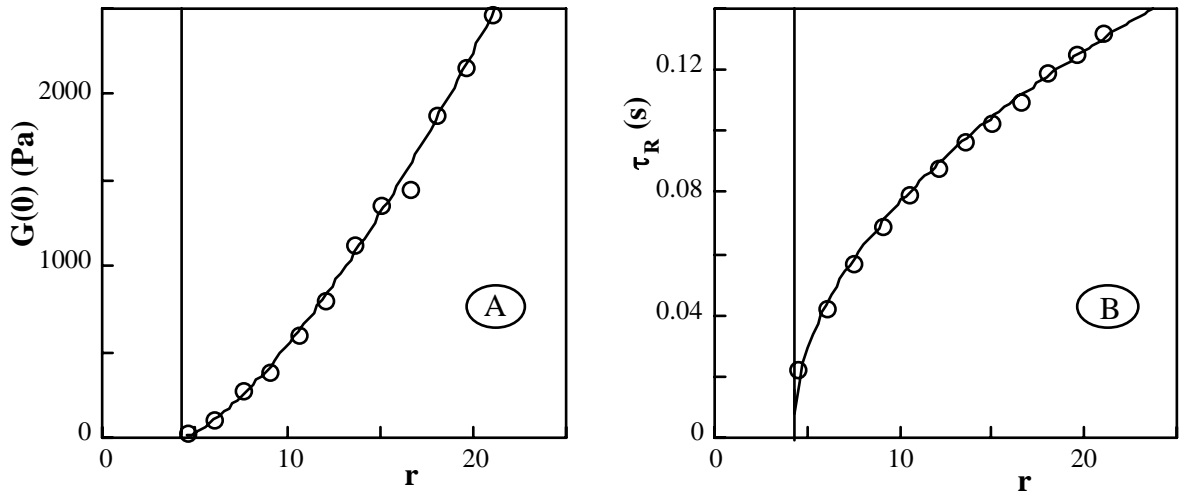

Michel et al Figure 9

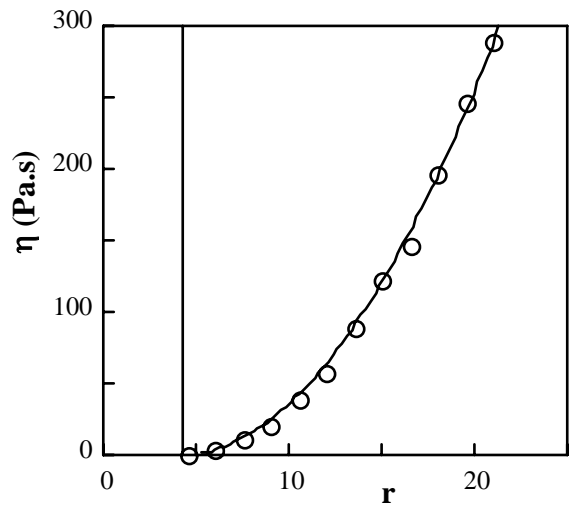

Michel et al Figure 10

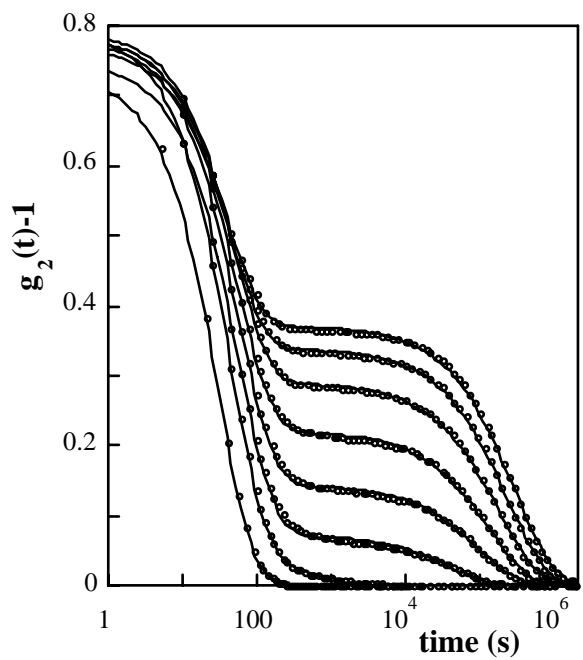

Michel et al   Figure 11

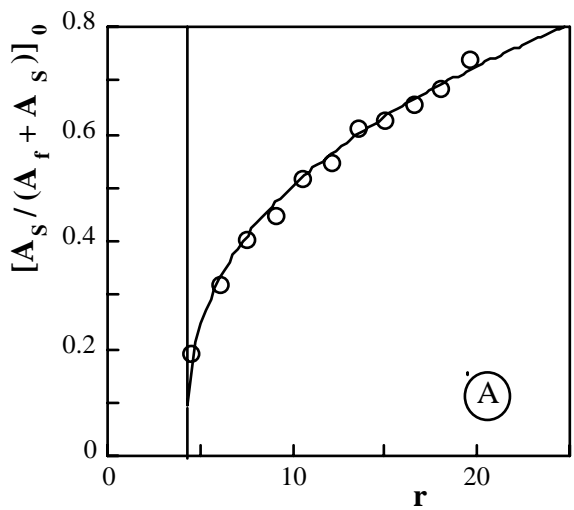
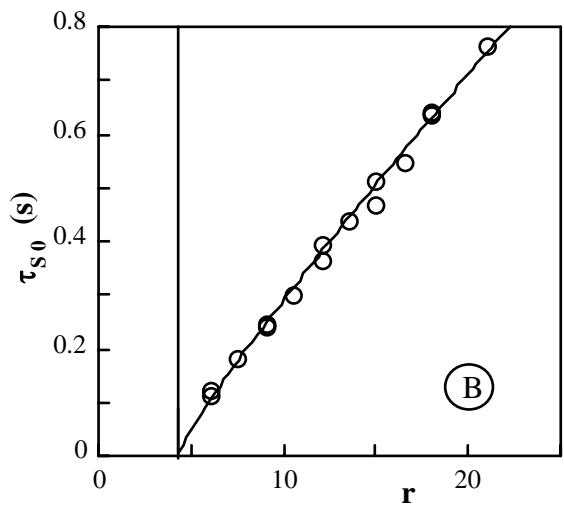

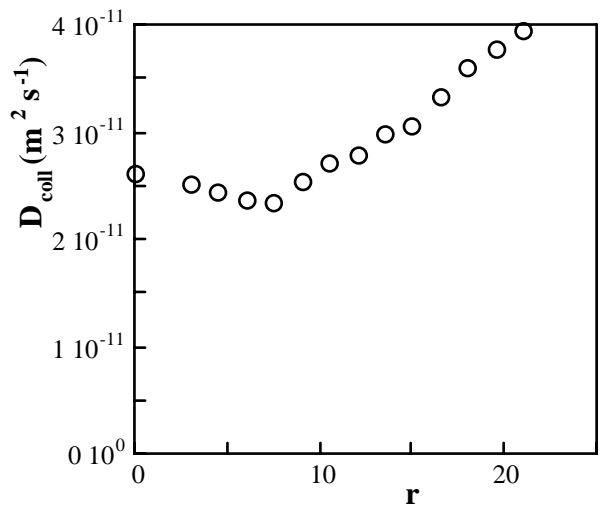

Michel et al Figure 13

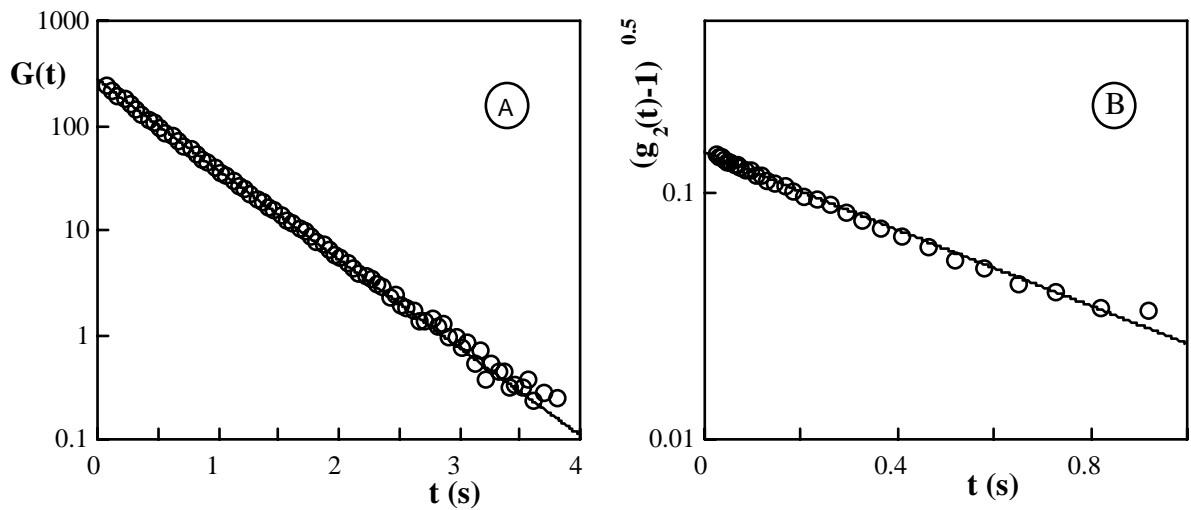